# Thermal stability and Thermal conductivity of pillared black phosphorene hybrid nanostructures:A molecular dynamics Study

Jiaxin Wu[1], Suxin Chen[1], Changpeng Chen[1,2,*] Yihan Ma[1], Yongshang Tian[3]

1. School of Science, Wuhan University of Technology, Wuhan 430070, P. R. China
2. Research Center of Materials Genome Engineering, Wuhan University of Technology, Wuhan 430070, P. R. China
3. Henan Key Laboratory of Utilization of Non-Metallic Mineral in the South of Henan, Xinyang Normal University, Xinyang 464000, P. R. China

## Abstract

In this study, a newly pillared black phosphorene hybrid nanostructures (PBP) is constructed with its thermal stability as well as thermal conductivity are studied by performing equilibrium molecular dynamics simulation (EMD). The results show that the curvature of the nanotube and the intense thermal vibration of the atoms are responsible for the failure of PBP at finite temperature simultaneously. The calculation of normalized HCACF and running thermal conductivities (RTC) as a function of correlation time demonstrate that thermal conductivities decreases with the temperature significantly, which is attributed by phonon coupling and scattering from phonon spectrum analyzing. Our result indicates that the thermal conductance of pillared black phosphorene is comparatively low as to that of pillared graphene, black phosphorous, and black phosphorus nanotubes, which may be promising for applications of some specific low dimensional nanostructure based thermal and nanoelectronic devices.

# Introduction

Since the experimental realization of phosphorus preparation, the research of this novel two-dimensional (2D) materials has attracted extensive attention. Black phosphorus(BP) has sp3 hybridized bond, which is the most stable structure among the phosphorus allotropes[1][2][3][4]. It has attracted tremendous research interest since it was successfully exfoliated from black phosphorus[5][6][7].In contrast to graphene, monolayer phosphorene has a direct band gap of 1.51 eV ,which has fascinating electronic properties including high carrier and a large on-off current ratio[7][8][9][10]. These superior electronic properties provide phosphorene a wider application future in many nanoelectronic and optoelectronic applications.

Apart from the special electronic and optical properties, the thermal properties of nanomaterials have drawn extensive attention for their unique features which distinct them from their counterpart in macroscale[11]. As for nanoscale devices with high power density, thermal management is the key point, because Joule heat needs to be dissipated rapidly by means of materials with high thermal conductivity. Furthermore, materials with low thermal conductivity are also useful to achieve high efficiency in electron transfer. As a novel semiconductor, phosphorene has been considered as a potential thermoelectric material

[12][13], thus an in-depth study on phosphorene is worthwhile. The thermal conductivity of black phosphorus is mainly controlled by phonons. Theoretically, the thermal conductivity of SLBP was mainly computed using the Boltzmann transport equation (BTE) approach where phonon-phonon scatterings are described by anharmonic lattice dynamics[14][15][16][17][18].

Recent studies have shown that the hybrid systems of two-dimensional materials can introduce additional structural diversities to further enrich the property and application space of 2D materials, at the same time provide an alternative to facilitate their fabrication and design processes, make up for the defects of a single structure, thus achieving ideal electrical and thermal properties. Pillared-graphene, which is the first reported effort to construct this nanostructure was presented by Kondo et al.[19] They presented a synthesis method that involved growing graphene multi-layers (or graphite) on the ends of vertically aligned multi-walled carbon nanotubes via chemical vapor deposition. After that, scholars have carried out a lot of research on pillared-graphene. G . C. Loh et al[20] which studied the effects of mechanical strain on the interfacial phonon dynamics in the pillared-graphene. A.Lakshmanan et al[21] calculated the thermal conductivity of pillared-graphene-epoxy, which indicated that the intercalation of the epoxy matrix can also promote a weaker dependence of conductivity on thermal variations.

Kun Cai et al [22] investigated the thermal stability of black phosphorus nanotube，which showed the tube should have a higher radius or should work at a lower temperature to achieve a stable BPNT with high robustness .

Referring to those previous work, we constructed a newly hybrid BP-BPNT junction (pillared black phosphorene) to investigate the thermal stability and thermal conductivity of the pillared black phosphorene by the equilibrium molecular dynamics simulation (EMD). The effects of temperature and structure are studied. In order to understand the underlying mechanisms, a detailed phonon spectra analysis is conducted in the latter part of this paper.

## Modeling

The junction linking the sheets and tubes can adopt many different geometries contingent on the chirality and tube radius. Here, we constructed three kinds of pillared black phosphorene (PBP) with different pipe diameter. The atomistic model for the PBP(pillared black phosphorene), which is illustrated in Fig.1, consists of armchair BPTN (A-BPTN) which is arranged in a regular hexagonal pattern with the chirality (7, 8)、(10,8)、(15,8) and the length of 12.87Angstroms (Å)、13.46Angstroms (Å) and 13.45Angstroms (Å). In order to make the BPTN and BP(Black phosphorene ) have equal lattice, we set the center of the BPTN coincides with the BP and the dimensions of BP used in the

PBP structure are approximately $79\times 74$, $75\times 75$ and $72\times 79$, respectively. The junctions are created by applying density function theory to freely connect and relax BPTNs with the defected BPs.

The thermal conductance is predicted by the Green-Kubo[24], which can enable the approach of equilibrium molecular dynamics to use just one single simulation to estimate the properties of tensorial nature. The Green-Kubo relation for thermal conductivity reads as [25][26][27]:

$$\kappa_{\mu\nu}(t) = \frac{V}{k_B T^2} \int_0^t dt' C_{\mu\nu}(t') \quad (1)$$

where $\kappa_{\mu\nu}(t)(\mu,\nu=x,y,z)$ represents the thermal conductivity tensor, $t'$ is the correlation time, $k_B$ is the Boltzmann's constant, T is the absolute temperature, and V is the volume of the simulation cell and the heat current autocorrelation function (HCACF) $C_{\mu\nu}(t)$ is defined as:

$$C_{\mu\nu}(t) = \langle J_\mu(0) J_\nu(t) \rangle \quad (2)$$

where J(t) is the microscopic heat current, which is defined as the time derivative of the sum of the moments of the site energies $E_i$ of the particles in the system[27][28]

$$J = \frac{1}{V} \frac{d}{dt} \sum_i r_i E_i \quad (3)$$

$$E_i = \frac{1}{2} m_i v_i^2 + U_i \quad (4)$$

Here $m_i$, $v_i$, and $U_i$ are the mass, velocity, and potential energy of particle i, respectively.

In the simulation, the periodic boundary conditions are applied in the plane expansion direction, the free boundary conditions are applied to another direction. Due to the SW potential developed by Jiang[29] which can perfect compute efficiency with reasonable accuracy for nonlinear behavior of a BP nanoscale structure, we choose it to describe the interaction among atom in the structure [30].

During the simulation, the initial system is firstly equilibrated by the constant volume and temperature ensemble (NVT) for 2 million time steps, the temperature increases from 10 K to a specified temperature with the gradient 10K. Then the system is switched to the constant volume and energy (NVE) for 20 million time steps. The NVE simulations continued for additional 2 000 000 time steps, with the fluctuating heat current data being collected every 10 ps. The number of heat current auto-correlation function (HCACF) is 100. In the time integration, we set the time step 0.1 fs.

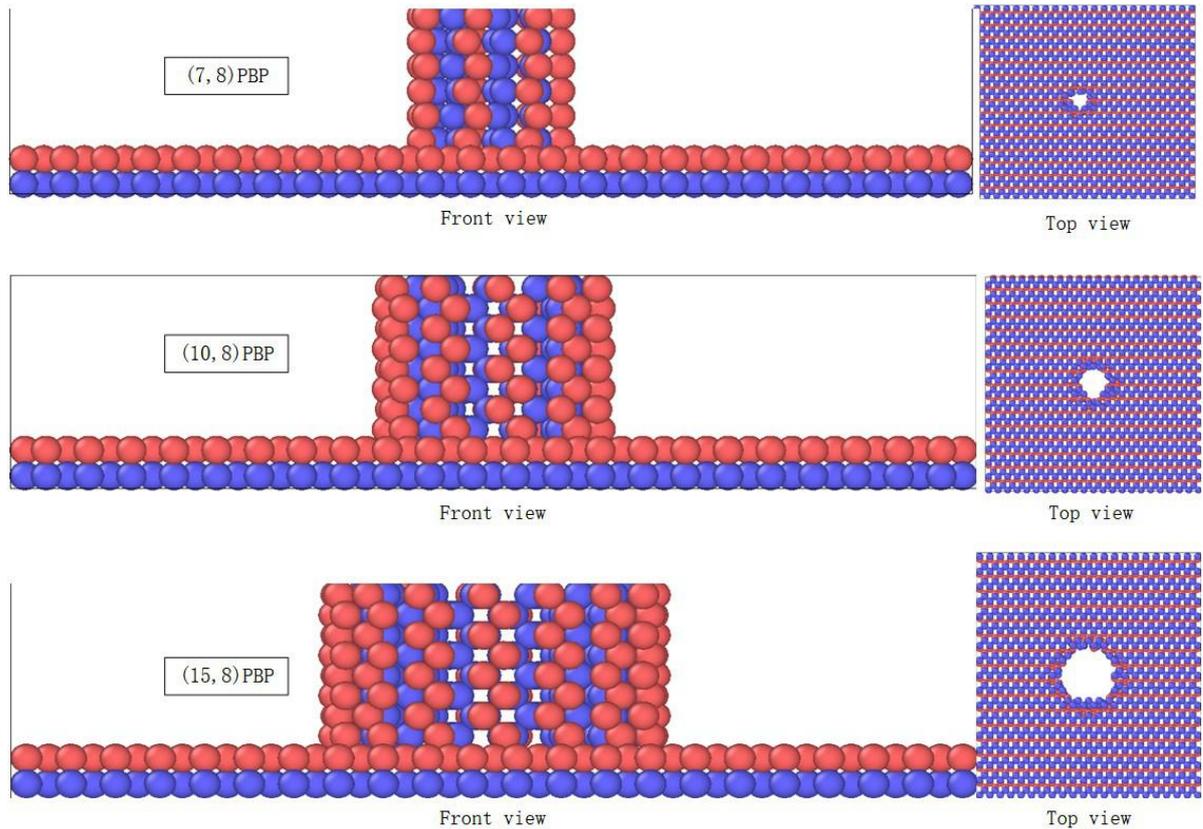

Fig.1 Atomic structure of three kinds of pillared black phosphorene

# 3.Results and discussion

## 3.1 Thermal stability

After a series of tests, we find that the model of PBP with the chirality (7, 8) configuration is damaged at 190K. The PBP with the chirality (10, 8) collapses at 210K within a few ps, and PBP with the chirality (15, 8) is stable till up to 230K. Figure 2 shows how the model changes during the simulation. It is phosphine nanotubes that for the geometric mapping which the inner layer of the tube that becomes shorter and the distance between the inner atoms and the adjacent atoms is shortened. The shorter the distance between the nonbonded atoms in the inner layer, the stronger the gravitation (VDW interaction) will be at the same time. The P–P bond on the outer tube is subject to tension due to the bending of the tube. In this case, the tube cannot maintain a stable structure, and the thermal vibration of atoms on the tube is also the major reason for tube destruction[22] .However, for the pillared black phosphorene, the surface structure has certain constraints on the expansion and the contraction of the nanotube which result in the stable existence of 7-PBP and 10-PBP at a higher temperature than the corresponding tube as shown in Fig.2 (a) and (b). With the increase of the diameter of the tube, the contact aperture between the plane and the nanotube increases. The thermal vibration, the expansion and contraction of the nanotube result in the decrease of the plane stability as shown in Fig.2 (c). Therefore, the 15-PBP structure collapse at a temperature of 230K.

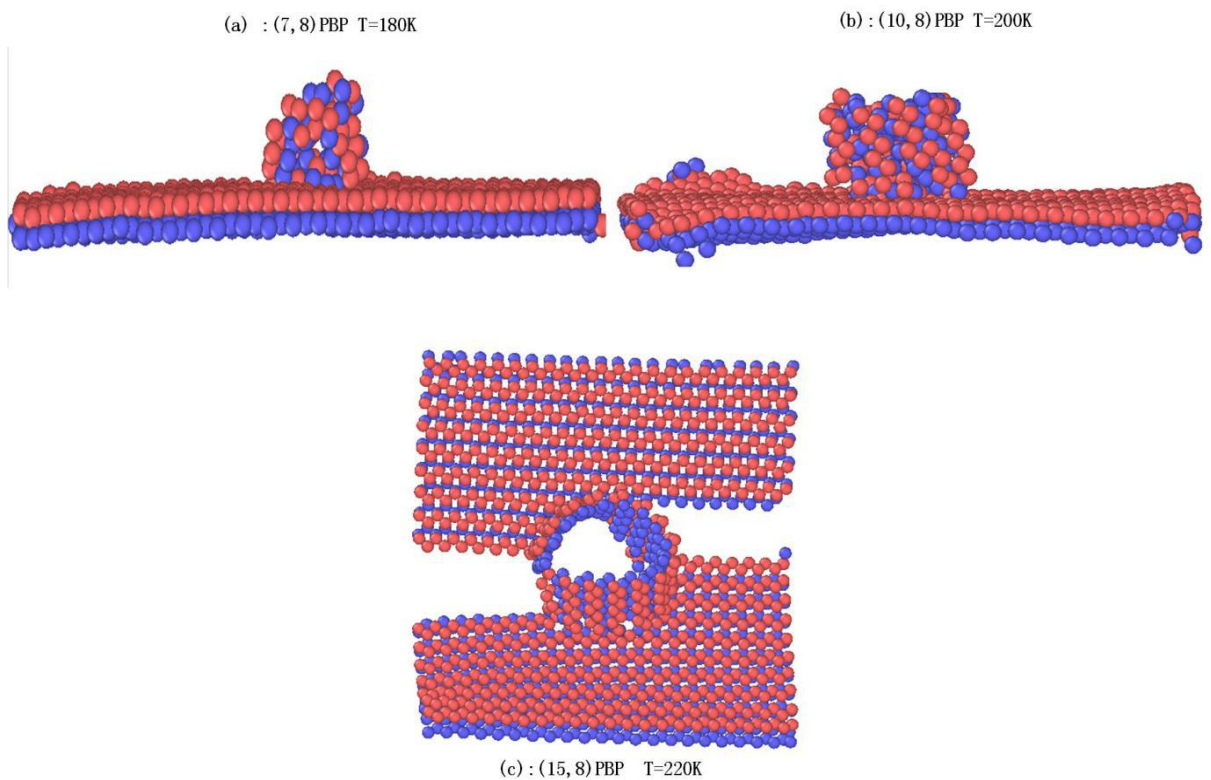

Fig.2 The snapshots of related models before model damage.

## 3.2 Thermal conductivity

### 3.2.1 Effects of temperature

We start by discussing the results for bulk pillared black phosphorene. The converged thermal conductivities of 7-PBP obtained by using different temperatures: T=10K,50K,100K,150K and 180K and the normalized HCACF are presented in Fig.3. Running thermal conductivities (RTC) as a function of correlation time for 7PBP changes at different temperatures. The thinner and thicker lines represent the results of independent simulations with different initial velocities, respectively. The running thermal conductivity (RTC) can vary from simulation to simulation and the variation increases with increasing correlation time. The figure of the normalized HCACF shows the heat current auto-correlation function fluctuates in a certain range, which means that the variation in the HCACF does not decay with increasing correlation time. This is a general property of time-correlation functions and transport coefficient in MD simulations[31] . The figure of running thermal conductivities shows the variety of thermal conductivity converging with the increase of correlation. The finite size effect of simulated region is an intrinsic property in EMD simulation, thus we can take the average values of thermal conductivity for different simulation temperatures as independent simulation results, in addition, we can obtain an average value of the corresponding error estimate. In this way we get the result that, with the increase of temperature, the thermal conductivity decreases. The thermal conductivity has a large decrease range at 10K-50K, however at 50K-180K, the decreasing trend of thermal conductivity becomes slower. The 10-PBP with 5 simulation temperatures, T=10K, 50K, 100K, 150K, and 200K, are shown in Fig.4, the relationship between thermal conductivity and temperature is presented in Fig. 6. It shows the running thermal conductivity (RTC) and HCACF converge with increasing correlation time, similar to 7PBP, the thermal conductivity decreases greatly at 10K-50K.

Finally, the thermal conductivity of 15-PBP with 5 simulation temperatures (T=10K, 50K, 100K, 150K, 220K) are shown in Fig.5, with the corresponding converged values presented in Fig.6. Due to the shorter correlation time required for HCACF converges, the total simulation time required for obtaining smooth curves of the RTC is shorter than that for 7-PBP, being about $N_s t_s = 1$ ps. But the change trend of thermal conductivity with temperature is similar to that of 7-PBP.

The thermal conductivity of phonon is:

$$\kappa = \frac{1}{3} C_V V \lambda \tag{5}$$

where $C_V$ refers to the lattice specific heat, V is the mean group velocity of phonons and $\lambda$ is the mean free path of phonon. When the temperature is relatively low, the thermal conductivity of the material is that of the phonons'. The factors that determine the thermal conductivity involves the heat capacity, the average velocity of phonon and the average free path of phonon, which are represented by $C_V$, V and λ, respectively. According to Dubai model, the specific heat approximation of phonons does not affected by temperature when the latter is relatively high. The temperature dependence of thermal conductivity is determined by the mean free path of phonons[23]. The number of phonons is:

$$n_{(q)} \approx \frac{\kappa_B T}{\hbar \omega_{(q)}} \qquad (6)$$

It can be seen that the number of phonons is directly proportional to the temperature, the number of phonons participating in the inversion process is proportional to the temperature, and the collision probability is proportional to the temperature as well. $\frac{1}{\tau} \propto T \rightarrow \lambda \propto \frac{1}{T}$. The inverse ratio of thermal conductivity to temperature can be concluded from Eq. (5) and Eq. (6). Hence the thermal conductivity decreases with the increase of temperature.

From Fig.6 one can see the thermal conductivity decreases with the increases of temperature, which can be explained as follows: the number of phonon decreases under relatively low temperature, which leads to less phonons scattering on the interface. When the temperature rises, more high frequency phonons are activated, making extra contribution to the thermal conducting. In the meantime, more phonon scatter on the inelastic interface under higher temperature. That leads to model deformation, it also increases in both the anharmonicity of atomic interactions and the phonon scattering of interfaces. Since the thermal conductivity is inversely proportional to temperature, the thermal conductivity decreases greatly at 10K-50K and then with the increase of temperature, the decreasing trend of thermal conductivity becomes slower.

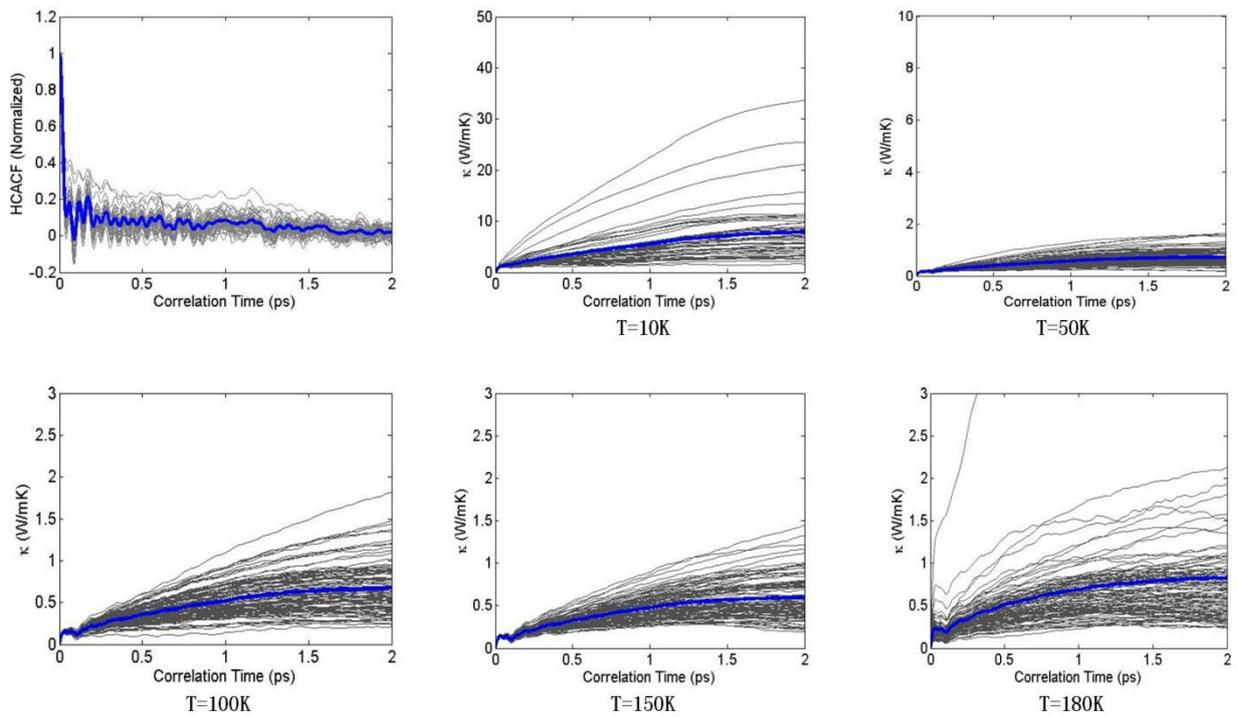

Fig.3 (a) The normalized HCACF and running thermal conductivities as a function of correlation time for 7-PBP with different simulation temperature . The thinner (and lighter) and the thicker (and darker) lines represent the results of independent simulations with different initial velocities and the ensemble average over the independent simulations, respectively.

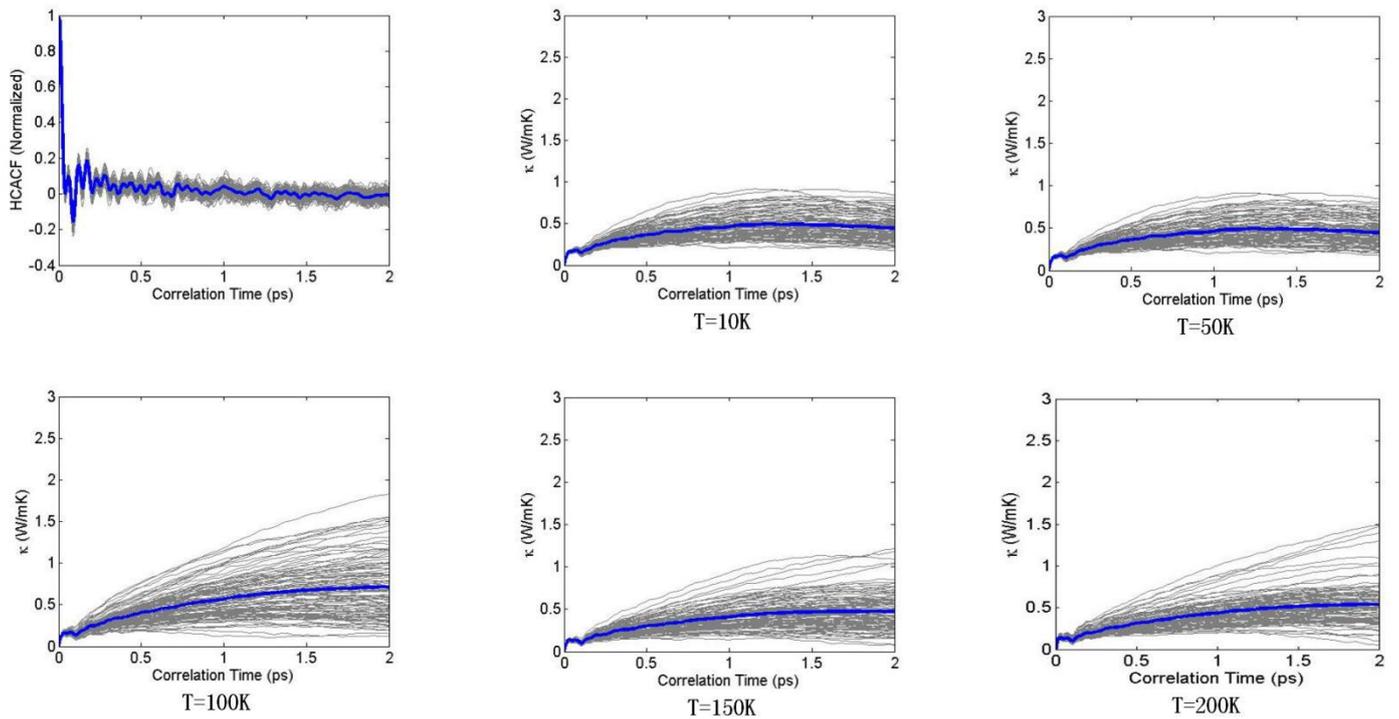

Fig. 4 Same as Fig. 3, but for 10-PBP.

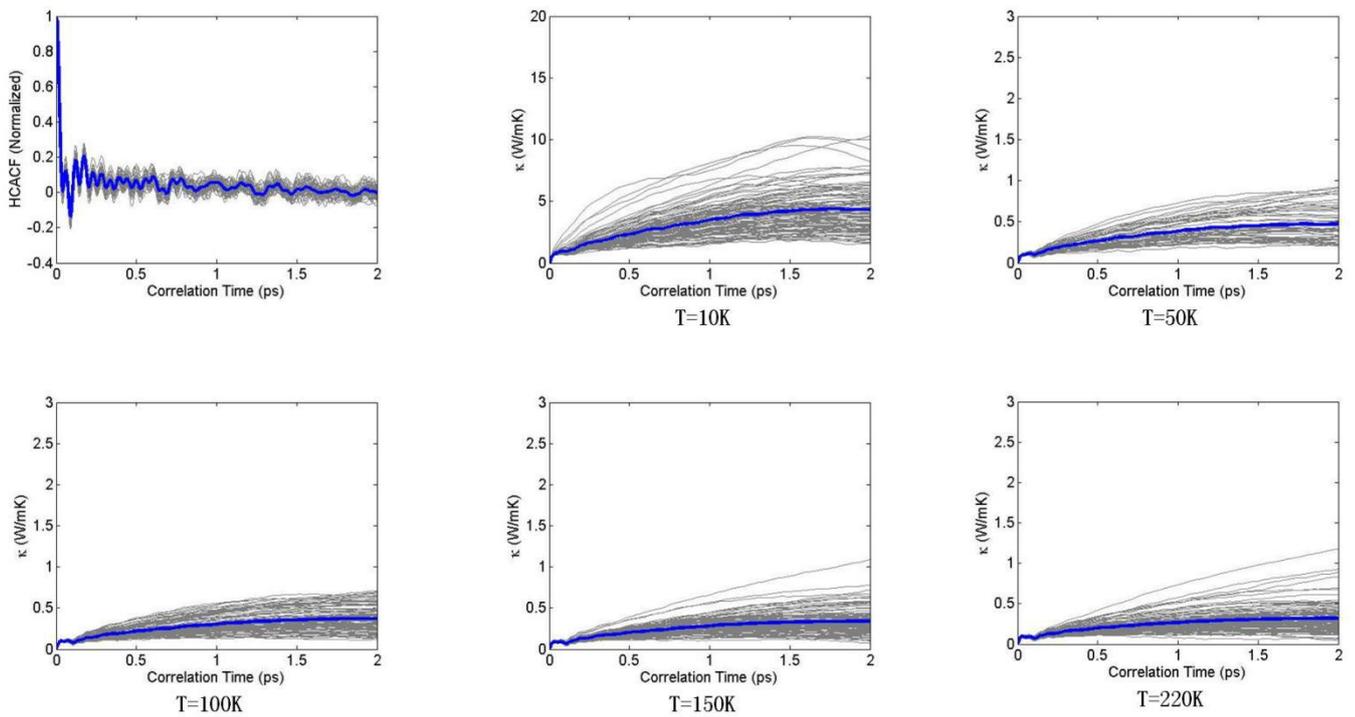

Fig. 5 Same as Fig. 3, but for 15-PBP.

### 3.2.2 Effects of structure

In order to analyze the relationship between thermal conductance and the different pillared black phosphorene structure, we plot the thermal conductivity with different structures in Fig.6. It shows that with the increase of temperature, the thermal conductivity of 7PBP is higher, the thermal conductivity of 10PBP and 15PBP cross at 10K-50K and the decreasing trend of thermal conductivity of 10PBP and 15PBP becomes slower when the temperature is higher than 50K. It is clear that the structure has a significant effect on thermal conductivity of pillared black phosphorene. With the increase of the contact aperture between the plane and the nanotube, the scattering at the phonon interface becomes more intense, which leads the thermal conductivity decreases.

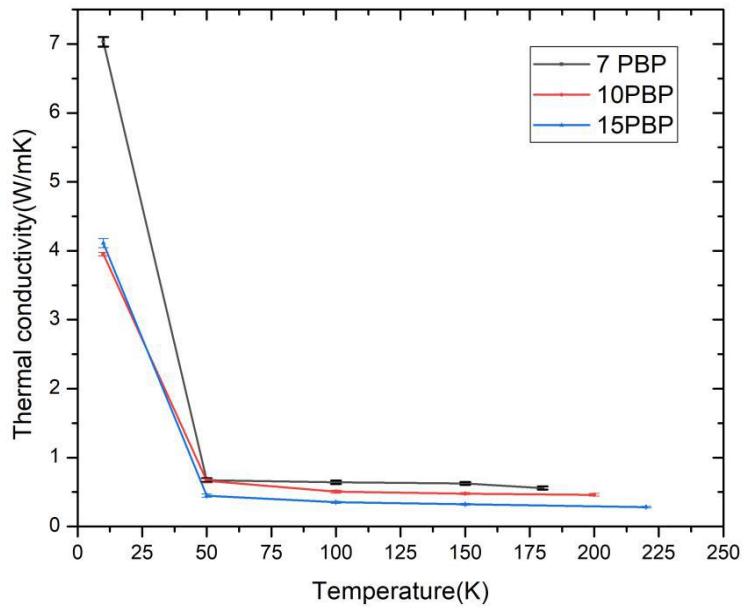

Fig.6　Temperature dependence of the thermal conductivities in the 7-PBP（black）,10-PBP(red) and 15-PBP(blue) of phosphorene for the three models

Phonon spectral analysis is further conducted to understand simulation results. The phonon density of states (PDOS) can be found through the

Fourier transform of the velocity autocorrelation function (VACF), and velocity autocorrelation function can be found by time-averaging[32]:

$$C_v(t) = \langle \vec{v}_i(0) \times \vec{v}_i(t) \rangle$$
$$= \frac{1}{N}\sum_{i=1}^{N}\frac{1}{t}\sum_{n=1}^{t}(\vec{v}_i(t_0) \times \vec{v}_i(t_0 + n\Delta t)) \quad (7)$$

where N is the total number of atoms and t is the total number of time steps. Finally, the phonon spectrum can be determined by the Fourier transform of the VACF[33]:

$$G(\omega) = \frac{1}{\sqrt{2\pi}}\int e^{i\omega t}\frac{\langle \vec{v}(t)\vec{v}(0)\rangle}{\langle \vec{v}(0)\vec{v}(0)\rangle}dt \quad (8)$$

where $\vec{v}(t)$ is velocity vector of an atom, and $\vec{v}(0)$ is the initial velocity vector. In spite of high quality but low frequency of the system, the sampling interval is set to 0.01ps. Figure 7(a) shows the velocity autocorrelation function (VACF) of the different pillared black phosphorene structure. We observe the velocity autocorrelation function (VACF) converging when the time is more than 5 ps, however, the amplitude of 15pbp is smaller, the amplitude of 10pbp and 7pbp are crossed. Figure 7(b) shows the calculation results of PDOS. For a given frequency, higher PDOS means that more states are occupied by for a phonon. Whereas zero PDOS means no phonon exists in the system. One of the crucial factors in determining the thermal resistance is the number of phonon states. If the phonon population for a certain x is very low, the energy propagation by phonons of that wave vector will be highly restricted. By examining the PDOS as shown in Fig.7(b), we observe remarkable differences that would account for the vast disparity in the thermal conductivities of the different pillared black phosphorene .Three peaks exist at the frequency of 0THz ,12THz and 32THz for PDOS of the 15-PBP(blue line), which the contribution of low frequency phonons is larger .7-PBP (black line)and 10-PBP (red line)have two higher peaks at the frequency of 13THz and 32THz ,however 7-PBP has a low peak at 5THz. 7-PBP and 10-PBP high frequency phonons contribute a lot. According to phonons aerodynamic model and elastic phonons model, heat transfer mainly depends on phonons vibration and existence of interface leads to phonon scattering. At low temperatures, fewer phonons, especially those with high frequencies are excited to participate in the thermal transport. 15PBP has lower frequency phonons participate in the thermal transport, however it has large phonon interface scattering. 7-PBP has a low peak at low frequency, hence thermal conductivity of 7-PBP is higher, and the thermal conductivity of 15-PBP is higher than 10 -PBP at 10K . As temperature increases, more high frequency phonons are excited, which make extra contribution to the thermal transport. Moreover, higher temperatures also bring about increased inelastic interface phonon scattering, which also leads to an increases in both the anharmonicity of atomic interactions and the phonon transmission coefficients of interfaces. Hence thermal conductivity of 7-PBP is higher than 10-PBP and thermal conductivity of 10-PBP is higher than 15-PBP at 50K-220K.

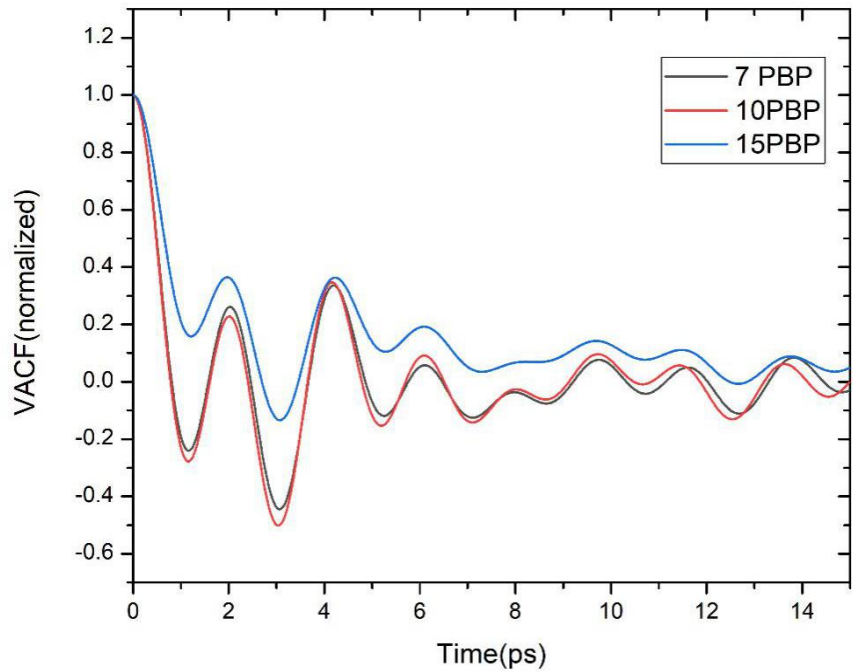

(a)

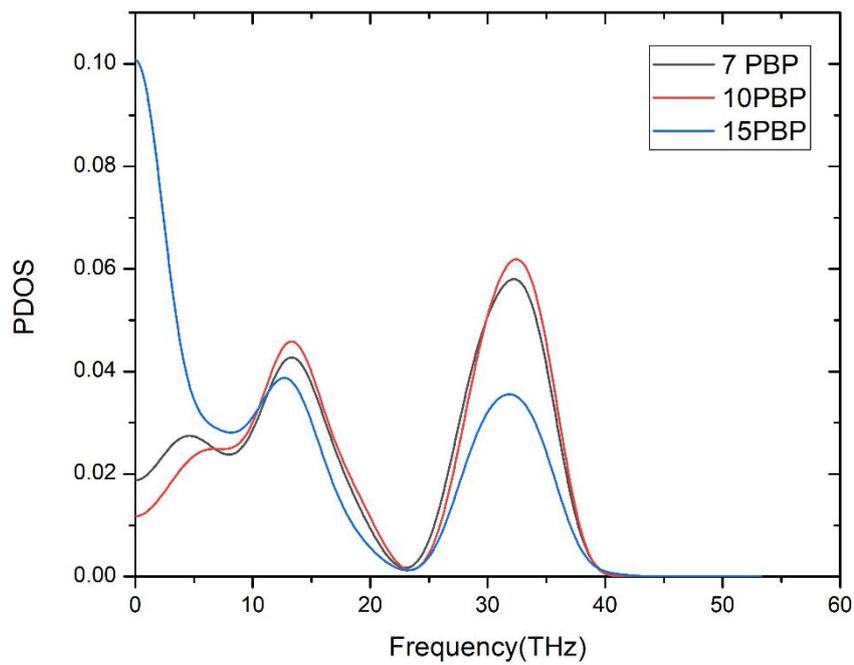

(b)

Fig.7 (a) Velocity autocorrelation functions of 7-PBP (black) , 10-PBP(red) and 15-PBP (blue). (b) The phonon DOS of the 7-PBP (black) , 10-PBP(red) and 15-PBP (blue)

We can find by comparison, the thermal conductivity decreases with the increase of nanotube diameter for the pillared black phosphorene. The thermal conductivity of pillared black phosphorene is 0.2821 ± 0.014W/mk～7.0323±0.69W/mk in this work, which is smaller than the that of pillared graphene(14.438W/mk ～ 392W/mk)in Ref.[34][35], armchair black phosphorus nanotubes(10W/mk ～ 24W/mk ) in Ref.[33] and black phosphorous (9.89W/mk ～ 110.7W/mk) in Ref.[36][37][38]. It can be used as a great thermal insulation material and has a wider application future.

## Conclusion

We constructed three kinds of pillared black phosphorene (PBP) with different pipe diameter, which are pillared black phosphorene with the chirality (7, 8) (7-PBP), pillared black phosphorene with the chirality (10, 8) (7-PBP) and pillared black phosphorene with the chirality (15, 8) (7-PBP). To find the effect of temperature on the thermal stability of the PBP at a canonical NVT ensemble, a series of MD simulations are presented ,in the simulation , we find that the 7-PBP model with the $79 \times 74$ dimension is damaged at 190K ,the 10-PBP with the $75 \times 75$ dimension collapses at 210K, and 15-PBP with the $72 \times 79$ dimension is stable below 230K. There are three mainly reasons for the failure of PBP at specified temperature:(1) the thermal stability of phosphorene nanotubes, (2) the thermal vibration of atoms, (3) the plane stability. Then We have studied the thermal transport of PBP using molecular dynamics simulations with SW potential. The normalized HCACF and running thermal conductivities (RTC) as a function of correlation time for PBP with different temperature are obtained. We analyze the effects of temperature and three different structures on the thermal conductivity of the PBP

For further explanation, we calculate the velocity autocorrelation function (VACF) and the phonon density of states (PDOS)of the different pillared black phosphorene structure. We explain the change of thermal conductance of pillared black phosphorene by phonon spectrum analyzing, phonon coupling and scattering. This work is helpful to understand the thermal transport behaviors in pillared black phosphorene and thermal stability of the pillared black phosphorene. It is helpful to promote their potential applications and enrich the knowledge of their research.

## Acknowledgments

The authors would like to acknowledge the support by 2018 China National College students' Innovative and Entrepreneurial Training Program Funding Projects and the National Natural Science Foundation of China (Project No. 61177076)

## References

[1] Jamieson J C 1963 Science 139 1291–2

[2] Kikegawa T and Iwasaki H 1983 Acta Cryst. B 39 158–64

[3] Akahama Y, Kobayashi M and Kawamura H 1999 Phys. Rev. B 59 8520–5

[4] Akahama Y, Kawamura H, Carlson S, Le Bihan T and Häusermann D. Phys. Rev. B 61 3139–42

[5] Liu H, Neal A T, Zhu Z, Luo Z, Xu X, Tomanek D and Ye P D 2014 ACS Nano 8 4033

[6] Reich E S 2014 Nature 506 19

[7] Li L, Yu Y, Ye G J, Ge Q, Ou X, Wu H, Feng D, Chen X H and Zhang Y 2014 Nat. Nanotechnol. 9 372

[8] Qiao J, Kong X, Hu Z X, Yang F and Ji W 2014 Nat. Commun.5 4475

[9] F. Xia, H. Wang, and Y. Jia, Nat. Commun. 5, 4458 (2014).

[10] Zhu Z , Chen C , Liu J , et al. Russian Journal of Physical Chemistry A, 2018, 92(1):132-139.

[11] N. Yang, X. Xu, G. Zhang, and B. Li, AIP Advances 2, 041410 (2012).

[12] R. Fei, A. Faghaninia, R. Soklaski, J.-A. Yan, C. Lo, and L. Yang, Nano Letters 14, 6393 (2014).

[13] J. Zhang, H. J. Liu, L. Cheng, J. Wei, J. H. Liang, D. D. Fan, J. Shi, X. F. Tang, and Q. J. Zhang, Sci. Rep. 4 (2014).

[14] L. Zhu, G. Zhang, and B. Li, Phys. Rev. B 90, 214302 (2014).

[15] Z. Y. Ong, Y. Cai, G. Zhang, and Y. W. Zhang, Journal of Physical Chemistry C 118, 25272 (2014).

[16] G. Qin, Q. B. Yan, Z. Qin, S. Y. Yue, M. Hu, and G. Su, Physical Chemistry Chemical Physics 17, 4854(2015).

[17] A. Jain and A. J. H. Mcgaughey, Scientific Reports 5,8501 (2015).

[18] G. Qin, X. Zhang, S.-Y. Yue, Z. Qin, H. Wang, Y. Han,and M. Hu, Phys. Rev. B 94, 165445 (2016).

[19] D. Kondo, S. Sato, and Y. Awano, Appl. Phys. Express 1, 074003 (2008).

[20] Loh G C , Teo E H T , Tay B K . Journal of Applied Physics, 2012, 111(1):6840.

[21] Lakshmanan A , Srivastava S , Ramazani A , et al.. Applied Physics Letters, 2018, 112(15):151902.

[22] Cai K, Wan J , Wei N , et al . Nanotechnology, 27(23):235703 (2016).

[23] Wu J , Wen H , Shi H , et al. Superlattices and Microstructures, 2019.

[24] S. Lepri, R. Livi, and A. Politi, Phys. Rep. 377, 1 (2003).

[25] M. S. Green, J. Chem. Phys.22, 398 (1954).

[26] R. Kubo, J. Phys. Soc. Jpn. 12, 570 (1957).

[27] D. A. McQuarrie, Statistical Mechanics, (University Science Books, Sausalito, 2000).

[28] Fan Z , Pereira L F C , Wang H Q , et al. Physical Review B, 2015, 92(9).

[29] Jiang J W .Nanotechnology .26 315706 (2015)

[30] Jiang J W, Rabczuk T and Park H S. Nanoscale 7 6059–68（2015）

[31] J. M. Haile, Wiley-Interscience, New York, (1992).

[32] Ng T Y, Yeo J , Liu Z . International Journal of Mechanics & Materials in Design, 9(2):105-114. (2013)

[33] Hao F , Liao X , Xiao H , et al. Nanotechnology, 2016, 27(15):155703.

[34] Varshney V , Patnaik S S , Roy A K , et al.. Acs Nano, 2010, 4(2):1153-61.

[35] Park J , Prakash V ., 2014, 116(1):7596.

[36] Y. Hong, J. Zhang, X. Huang, and X. C. Zeng, Nanoscale 7, 18716 (2015).

[37] Y. Y. Zhang, Q. X. Pei, J. W. Jiang, N. Wei, andY. W. Zhang, Nanoscale 8, 483 (2016).

[38] Ke X , Zheyong F , Jicheng Z , et al.. Modelling & Simulation in Materials Science & Engineering, 2018.